\documentclass{emulateapj}
\def\stacksymbols #1#2#3#4{\def\theguybelow{#2}
        \def\verticalposition{\lower#3pt}
        \def\spacingwithinsymbol{\baselineskip0pt\lineskip#4pt}
        \mathrel{\mathpalette\intermediary#1}}
\def\intermediary #1#2{\verticalposition\vbox{\spacingwithinsymbol
        \everycr={}\tabskip0pt
        \halign{$\mathsurround0pt#1\hfil##\hfil$\crcr#2\crcr
                \theguybelow\crcr}}}
\def\lta{\stacksymbols{<}{\sim}{2.5}{.2}}
\def\gta{\stacksymbols{>}{\sim}{3}{.5}}

\begin{document}

\title{CREATION OF X-RAY CAVITIES IN GALAXY CLUSTERS 
WITH COSMIC RAYS} 

\author{
William G. Mathews\footnotemark[1],
Fabrizio Brighenti{\footnotemark[1]$^,$\footnotemark[2]}
}

\footnotetext[1]{UCO/Lick Observatory,
Dept. of Astronomy and Astrophysics,
University of California, Santa Cruz, CA 95064}

\footnotetext[2]{Dipartimento di Astronomia,
Universit\`a di Bologna,
via Ranzani 1,
Bologna 40127, Italy}


\begin{abstract}
We describe how AGN-produced cosmic rays form 
large X-ray cavities and radio lobes in the hot diffuse gas in galaxy 
groups and clusters.
Cosmic rays are assumed to be produced in a small shocked 
region near the cavity center, such as at the working surface 
of a radio jet. 
The coupled equations for gasdynamics and cosmic ray diffusion
are solved with various assumptions about the diffusion 
coefficient.
To form large, long-lived cavities similar to those observed, 
the diffusion coefficient must not exceed $\kappa \sim 10^{28}$
cm$^2$ s$^{-1}$ in the hot gas, 
very similar to values required in models of 
cosmic ray diffusion in the Milky Way.
When $\kappa \lta 10^{28}$
cm$^2$ s$^{-1}$, cosmic rays are confined within 
the cavities for times comparable to the cavity buoyancy time, 
as implied by observations of X-ray cavities and their 
radio synchrotron emission. 
Collisions of proton cosmic rays with thermal plasma nuclei 
followed by $\pi^0$ decays can 
result in enhanced gamma ray emission from the cavity walls. 
\end{abstract}


\section{Introduction}

In recent years high resolution X-ray observations 
of the hot gas in galaxies, groups and clusters have 
revealed cavities or bubbles having sizes of a few kiloparsecs.
About 20 percent of all groups/clusters are currently known 
to contain these cavities 
(e.g. Bohringer et al. 1993;
McNamara et al. 2000; Birzan et al. 2004), 
but they may appear intermittently in all clusters. 
In previous computational studies
X-ray cavities have been produced 
by the introduction of ultrahot gas that expands, 
displacing the 
ambient hot gas in a purely gasdynamical manner
(e.g. Churazov et al. 2001, 2002;
Quilis et al. 2001;
Br\"uggen \& Kaiser 2002;
Br\"uggen 2003; Robinson et al. 2004;
Jones \& De Young 2005).
However, many cavities are filled with radio 
synchrotron and inverse 
Compton emission produced by relativistic electrons. 
Although it is often claimed that 
cavities are inflated by the pressure of cosmic rays, 
cavity formation 
by cosmic rays has not been investigated in detail. 
The interaction of cosmic rays with hot gas 
differs from ordinary gas dynamics 
because of the important role of cosmic ray diffusion. 
In the discussion below we describe the combined 
dynamics of gas and diffusing cosmic rays to determine 
those properties 
of cosmic rays -- energy density, luminosity, diffusion 
coefficient, etc. -- that are best suited to 
form X-ray cavities similar to those observed. 

It is widely assumed that mass accretion onto a central massive 
black hole (i.e. AGN) is the ultimate energy source 
that forms these cavities, but intermediate 
details are lacking. 
We assume here that cosmic rays are the agency that transmits 
AGN energy to the gas. 
Cosmic rays can be formed in shock waves  
at the tip (working surface) of a radio jet and 
this is supported by many observations. 
Cosmic rays may also be produced by shocks or 
other processes near the central black hole. 
In either case, as cosmic rays propagate into the hot 
ionized gas, they are scattered by irregularities in the 
magnetic field which, if frozen into the hot gas, convey  
cosmic ray momentum to the gas.

Once a low-density X-ray cavity forms at some 
radius in the atmosphere of hot cluster gas, 
its buoyancy transports it radially outward. 
Therefore, to be visible as isolated, coherent features, 
X-ray cavities must form 
during time scales $t_{lobe}$ that are considerably less 
than the buoyancy time $t_{buoy}$, which is in turn somewhat 
longer than the dynamical (freefall) time in the local 
galactic or cluster potential. 
However, cavities must not be formed explosively 
on very short time scales since this would create 
powerful shocks, strongly heating the gas in the cavity rims 
that is not observed.
Indeed, the gas in cavity rims is often the coolest 
gas in the entire hot gaseous atmosphere 
(e.g. Fabian 2001, 2002; Brighenti \& Mathews 2002). 

We present below a series of spherical, one-dimensional,
time-dependent 
calculations that describe how cosmic rays 
create X-ray cavities.
For simplicity we assume that the gas has 
uniform density and temperature prior to the introduction 
of cosmic rays. 
In this first study we concentrate on the formation 
and duration of 
cavities and ignore the additional complication of 
gravity which however is essential to understand 
the buoyant phase.

\section{Equations and Computational Procedure}

We assume that the cosmic rays (CRs) are 
produced in a shock located in a volume that is 
small compared to the cavity radius. 
The CR-gas interaction can be most simply
expressed with four Eulerian equations:
\begin{equation}
{ \partial \rho \over \partial t} 
+ {\bf \nabla}\cdot\rho{\bf u} = 0
\end{equation}
\begin{equation}
\rho \left( { \partial {\bf u} \over \partial t}
+ ({\bf u \cdot \nabla}){\bf u}\right) = 
- {\bf \nabla}(P + P_c)
\end{equation}
\begin{equation}
{\partial e \over \partial t}
+ {\bf \nabla \cdot u}e = - P({\bf \nabla\cdot u})
\end{equation}
\begin{equation}
{\partial e_c \over \partial t}
+ {\bf \nabla \cdot u}e_c = - P_c({\bf \nabla\cdot u})
+ {\bf \nabla\cdot}(\kappa{\bf \nabla}e_c)
\end{equation}
where artificial viscosity terms are suppressed. 
A similar set of equations has been discussed by 
Jones \& Kang (1990). 
Pressures and thermal energy densities in the plasma
and cosmic rays are related respectively by
$P = (\gamma -1)e$ and $P_c = (\gamma_c - 1)e_c$ 
where we assume $\gamma  = 5/3$ and $\gamma_c = 4/3$.
The cosmic ray dynamics are described by
$e_c$, the integrated energy density over the cosmic ray 
energy or momentum distribution,
$e_c \propto \int EN(E) dE \propto \int p^4 f(p)(1+p^2)^{-1/2} dp$. 
In general 
the CR diffusion coefficient $\kappa$ may depend on the 
particle momentum, $\kappa (p) \propto p^{0.5 - 0.6}$, as
suggested by CR models in the Milky Way
(Malkov 1999), but we do not consider this possibility here 
and regard both $e_c$ and $\kappa$ as mean values 
integrated over the CR spectrum.
The first three equations above describe the dynamical 
evolution of the hot gas influenced by the 
combined CR and gas pressure. 
The last equation describes the simultaneous 
gasdynamical advection and diffusion of the CR 
energy density.
A mass conservation equation for the CRs is unnecessary
because of their negligible rest mass.

Cosmic rays and hot plasma are coupled by mutual
interactions with a small (and otherwise dynamically insignificant)
magnetic field with topological irregularities 
that scatter and diffuse cosmic ray particles on small spatial scales. 
However, no magnetic terms need explicitly appear
in the equations provided magnetic stresses are small 
compared to $P$ and $P_c$.
This assumption is justified by the small magnetic fields 
observed in group/cluster gas, 
$\sim 1-10\mu$G (Govoni \& Feretti 2004), 
corresponding to energy densities $\sim B^2/8\pi \lta 10^{-11}$
erg cm$^{-3}$ that are generally much less than the thermal 
energy density in the hot gas.
Another implicit assumption in Equation (4) is that 
the (Alfven) velocity of the magnetic scatterers is small relative
to the hot gas 
(e.g. Drury \& Falle 1986).
This assumption is also reasonable for our purposes 
since the Alfven velocity 
$v_A = B/(4\pi \rho)^{1/2} = 2 n_e^{-1/2} B(\mu{\rm G})$
km s$^{-1}$ is generally very much less than 
the sound or flow speeds for magnetic 
fields and plasma electron densities of interest here. 

Our neglect of magnetic terms in Equations (1)-(4) 
also applies to the radio lobes.
Observations of inverse Compton (IC) X-rays
(i.e. upscattered cosmic microwave photons) from
many radio lobes in X-ray clusters
permit estimates of both the
cosmic ray and magnetic energy densities
(e.g. Longair 1994), independent of the
equipartition assumption (Burbidge 1959) which
may not be physically justified.
Croston et al. (2005) observed keV IC emission in
33 FRII sources in cluster cooling flows
and found magnetic fields within
0.3-1.3 of equipartition
assuming pure electron (e$^{\pm}$) cosmic rays.
The magnetic fields $B \sim 10\mu$G in the
radio lobes are comparable to
those estimated from Faraday depolarization throughout
all the hot gas
(e.g. Govoni and Feretti 2004) 
and are dynamically unimportant.
Croston et al. argue that cosmic ray protons
are not a dominant component
since equipartition can be achieved with electrons alone.
However, Croston et al. do not constrain the radio lobes to
be in pressure equilibrium with the local hot gas.
In their study of radio emission from
FRI lobes, Dunn \& Fabian (2004) assumed pressure
equilibrium with the surrounding gas and
found that the pressure of cosmic ray electrons
exceeded equipartition values in all sources, 
i.e. the particle energy density exceeds that of the 
magnetic field.
About half of the lobes they studied require
a dominant population of cosmic ray protons, 
in contrast with the results of Croston et al.
For our purposes here it is not necessary 
to specify the CR composition, either electrons or protons 
can dominate.
But we do assume that the total CR energy density is not 
substantially reduced during the cavity evolution time 
by losses due to synchrotron emission 
or interactions with ambient photons and thermal particles. 

In the model we adopt in Equations (1)-(4), we assume that 
the random fields that scatter CRs dominate over fields 
that are coherent on the scale of the radio lobe, consequently 
the diffusion coefficient $\kappa$ can be represented 
as a scalar rather than a tensor. 
This seems plausible because the shapes of X-ray cavities 
are to first order spherical, not highly elongated.  
The CR diffusion described by Equations (1)-(4) are appropriate 
for CR-scattering field irregularities (Alfven waves) 
that are coupled to the inertia of the plasma. 
It is less clear if the CRs diffuse in the same manner 
relative to the field 
inside the cavities/lobes where the plasma density may be 
very low, but we assume that they do.

\section{Lagrangian Calculations}

In this initial study of the combined dynamics of hot gas 
and cosmic rays we solve the one-dimensional spherical version of the 
equations above. 
While solutions can be found in either Eulerian or Lagrangian 
computational grids, we discuss the Lagrangian solutions here 
because they are conceptually simpler, require fewer 
computational adjustments, and show more dramatically 
how the gas is cavitated by the cosmic rays. 
The Lagrangian equations are
\begin{equation}
{\partial u \over \partial t} = 
{1 \over \rho} {\partial \over \partial r}(P + P_c + Q)
\end{equation}
\begin{equation}
u = {\partial r \over \partial t}
\end{equation}
\begin{equation}
\rho = {3 \over 4 \pi}{\partial m \over \partial (r^3)}
\end{equation}
\begin{equation}
{\partial \varepsilon \over \partial t}
= {(P + Q) \over \rho^2}{\partial \rho \over \partial t}
\end{equation}
and
\begin{equation}
{\partial e_c \over \partial t} 
= {\gamma_c e_c \over \rho} {\partial \rho \over \partial t}
+ {1 \over r^2}{\partial \over \partial r}
\left(r^2 \kappa {\partial e_c \over \partial r}\right).
\end{equation}

The first four equations are the usual Lagrangian 
gasdynamical equations 
where $m$ is the mass coordinate and $\varepsilon = e/\rho$ is 
the specific thermal energy. 
Differenced versions of these equations are solved 
explicitly at each timestep for time-advanced 
quantities -- $u$, $r$, $\rho$, $\varepsilon$ and $e_c$ -- 
in the same order as the equations above.
Equations (8) and (9) are solved term by term using
operator splitting. 
The diffusion term in Equation (9) is solved
with an implicit Crank-Nicolson difference scheme, 
stable (but not necessarily accurate) for any time step.
Although the Lagrangian grid necessarily follows the gas as it is 
pushed away from the source of cosmic rays, the cosmic rays 
continue to fill the central cavity left behind. 
Consequently, the cosmic ray diffusion term in 
Equation (9) is solved on a stationary  
Eulerian grid within the cavity and on 
a time-varying Lagrangian grid in the gas. 
Both Eulerian and Lagrangian grids are updated 
as necessary at each time step. 
The computational 
time step $\Delta t = a\min(\Delta t_c, \Delta t_d)$ 
is a fraction $a < 1$ of the minimum 
of the Courant time step $\Delta t_c = \Delta r/c_s$ 
(with sound speed $c_s = [(\gamma P + \gamma_c P_c)/\rho]^{1/2}$) 
and the time step for stability of 
{\it explicit} differencing of the diffusion term,
$\Delta t_d = (\Delta r)^2 /4 \kappa$. 
In zones where the gas is strongly compressed, 
$\Delta r$ becomes small, slowing the calculation. 
Therefore we remove zones by combining adjacent 
Lagrangian zones (conserving mass and energy) 
whenever $\Delta r$ drops below an $r$-dependent
minimum value. 

Since the hot gas is strongly compressed in a large 
region ahead of the diffusing cosmic rays,
the artificial viscosity $Q$ needs some modification 
to avoid excess dissipation. 
In addition to the standard Richtmeyer-von Neumann 
definition of $Q$ which is nonzero only in in compressing zones, 
we require that (1) $Q = 0$ if 
the gas gradient $\partial \log P / \partial \log r < 0.3$ across a zone 
and (2) $Q$ is not allowed to exceed the local gas pressure $P$. 

For these calculations 
we regard the CR diffusion coefficient $\kappa$ as an 
unknown parameter. 
In the absence of detailed information about the nature of the 
magnetic scattering, it is difficult
to derive $\kappa$ from first principles.
However, $\kappa$ can be estimated for the X-ray cavity problem 
from the global properties of observed cavities. 
As CRs diffuse away from a small source, their pressure gradient 
displaces the ambient hot plasma until a cavity 
of radius $r_{lobe}$ is formed after time $t_{lobe}$.
Since the CR diffusion coefficient
$\kappa$ has dimensions length$^2$/time,
a value $\kappa \approx r_{lobe}^2/t_{lobe}$ is required 
to establish a CR pressure gradient 
$\partial P_c / \partial r$ over distance $r_{lobe}$ 
in time $t_{lobe}$.
Localized, quasi-spherical X-ray cavities must be formed 
in a time $t_{lobe}$ that is 
significantly less than the buoyancy rise time $t_{buoy}$
of the cavity in the hot gas atmosphere of the galaxy 
group/cluster.  
(If $t_{lobe} \gta t_{buoy}$ then plume-like cavities would 
form and this is not generally observed.) 
Typically $t_{buoy}$ 
is several times larger than the dynamical 
(freefall) time $t_{dy}$ at the 
radius $r$ where the cavity is formed in the group/cluster potential.
Consequently, the diffusion coefficient (diffusivity) is 
of order $\kappa \approx 3.0 \times 10^{28}
r_{lobe,kpc}^2/(t_{lobe}/10^7~{\rm yrs})$ cm$^2$ s$^{-1}$ 
where $\sim 10^7$ yrs is a typical dynamical time within 
a few kpc of the centers of large elliptical galaxies. 
Therefore, a diffusivity $\kappa \sim 10^{30}$ cm$^2$ s$^{-1}$ 
in the cavity region is required to form cavities of size 5 -- 8 kpc.

However, in the more acceptable 
calculations described below a somewhat 
smaller $\kappa \sim 10^{28}$ cm$^2$ s$^{-1}$ is required in the 
initial gas. 
The reason for this is that  
the cavities must persist for times $t_{buoy} \sim 10^8$ yrs, 
and for these 
long-lived cavities the relevant length scale that determines 
$\kappa$ is not $r_{lobe}$ but 
the thickness of the cavity wall, $\lta 0.1r_{lobe}$.
Consequently, cavities last for times $\sim t_{buoy}$ 
only if the diffusivity in the undisturbed 
initial gas is $\kappa \lta 10^{28}$ 
cm$^2$ s$^{-1}$.
This is a remarkable result since $\kappa \sim 10^{28}$
cm$^2$ s$^{-1}$
is the value of the diffusivity most favored in studies
of CR diffusion in the Milky Way disk and halo
(Jones, 1979;
Strong \& Moskalenko 1998;
Snodin, et al. 2006).
This noteworthy and encouraging 
consistency suggests that X-ray cavities are indeed
formed by cosmic rays.
At CR particle velocities
$\langle v \rangle = 1.4\times 10^9 E_{Mev}^{1/2}$ cm s$^{-1}$,
the diffusivity 
$\kappa \approx \langle v \rangle \lambda \approx 10^{28}$
cm$^2$ s$^{-1}$
suggests a 
mean free path, $\lambda \approx 2/E_{Mev}^{1/2}$ pc, 
that is much less than astronomical scales of interest, 
although this simple kinetic model for the diffusion is 
almost certainly inadequate. 
In the exploratory calculations described here we begin 
by assuming a uniform diffusivity $\kappa$ throughout the 
cavity and gas.
However, the diffusion of cosmic rays 
is expected to depend on the density of 
frozen-in magnetic
scatterers which scales with the gas density. 
For this reason 
we also consider solutions in which
$\kappa$ varies inversely as a power of the gas density, 
$\kappa \propto \rho^{-q}$.

In solving Equation (9) 
it is necessary to solve the time-dependent diffusion equation 
\begin{equation}
{\partial e_c \over \partial t} =
{1 \over r^2}{\partial \over \partial r}
\left(r^2 \kappa {\partial e_c \over \partial r}\right)
\end{equation}
at each time step 
over the entire grid (cavity plus gas). 
For a sufficiently large computational region 
we can safely assume that the 
CR flux $r^2 \kappa \partial e_c / \partial r$ vanishes 
at the outer boundary and apply reflecting boundary 
conditions there. 
However, at the center of the flow 
we must include an appropriate source of CRs. 
We describe the CR flux in the innermost computational zone using 
the steady-state point-source solution of the equation above 
\begin{equation}
e_c(r) = { {\dot E}_c \over 4 \pi \kappa} {1 \over r}
\end{equation}
where ${\dot E}_c$ is the CR luminosity (ergs s$^{-1}$) 
at the origin.
Therefore, in solving Equation 10 
as $r \rightarrow 0$, we assume that the flux 
$r^2 \kappa \partial e_c / \partial r$ converges to 
$- {\dot E}_c / 4\pi$, as in the steady state point source 
solution. 
This approximation should be accurate close to the CR source.

If an X-ray cavity of radius 
$r_{lobe}$ is formed in time $t_{lobe}$, then 
the approximate 
energy required to form and fill the cavity with CRs is
\begin{equation}
E_{lobe} = {\gamma_c \over \gamma_c - 1} 
P_0V_{lobe} ~~~{\rm where}~~~V_{lobe} = {4 \over 3}\pi r_{lobe}^3,
\end{equation}
the CR luminosity is 
\begin{equation}
{\dot E}_c = E_{lobe}/t_{lobe},
\end{equation}
and the characteristic cavity diffusivity is 
\begin{equation}
\kappa_0 = r_{lobe}^2/t_{lobe}~~~{\rm and}~~~
\kappa = c_{\kappa} \kappa_0
\end{equation}
where $P_0$ is the pressure of the hot gas 
initially surrounding the CR source and $c_{\kappa}$ is a 
dimensionless coefficient. 
The energy $E_{lobe}$ 
includes the work done in displacing the gas 
$P_0V_{lobe}$ and the CR energy within the cavity at 
pressure equilibrium, $(\gamma_c - 1)^{-1}P_0V_{lobe}$, 
which is less accurate during the period of cavity formation 
$t \lta t_{lobe}$. 

Based on these dimensional considerations, 
we assume in the following calculations 
that the CR luminosity ${\dot E}_c$ is constant for 
$t < t_{lobe}$, and zero thereafter. 
To generate various types of solutions, we vary 
$c_{\kappa}$ but keep 
$t_{lobe} = 10^7$ yrs and $r_{lobe} = 6$ kpc fixed in 
all calculations. 
To avoid an unnecessary excess of parameters, we assume 
that all cavities are produced in a uniform gas of 
initial density $n_{e0} = 0.03$ cm$^{-3}$ and temperature 
$kT = 2$ keV ($2.32 \times 10^7$ K).
We assume throughout relativistic cosmic rays with
$\gamma_c = 4/3$.
With these assumptions 
the sound speed in the undisturbed gas is 
$c_{s0} = 726$ km s$^{-1}$, 
the initial gas pressure is $P_0 = 1.84 \times 10^{-10}$ dyne 
cm$^{-2}$, 
the CR luminosity is 
${\dot E}_c = 6.18 \times 10^{43}$ erg s$^{-1}$, 
the lobe energy is $E_{lobe} = 1.94 \times 10^{58}$ ergs,
and $\kappa_0 = 1.08\times 10^{30}$ cm$^2$ s$^{-1}$.

\section{Cavity with Constant $\kappa$}

We begin with a discussion of a solution of Equations 5--9 
in which the diffusion 
coefficient $\kappa = \kappa_0 = 1.08 \times 10^{30}$ cm$^2$ s$^{-1}$
is constant everywhere, in the cavity and throughout the gas 
i.e. $c_{\kappa} = 1$. 
We assume that there is enough residual plasma in 
the cavity to sustain magnetic fields and provide inertial 
resistance to the diffusing CRs. 
The presence of magnetic fields within observed cavities 
is obvious from the enhanced radio-synchrotron 
emission observed there.
As described above, the CRs are assumed to turn on at $t = 0$, 
maintain a constant luminosity ${\dot E}_c$, 
and turn off abruptly at $t_{lobe} = 10^7$ yrs.

Figure 1 shows the flow evolution at four 
representative times $t \le 1 \times 10^7$ yrs. 
The large cosmic ray pressure gradient within about 2 kpc 
dominates the initial formation of the cavity. 
The leading edge of the expanding gas  
is a region of compression 
where the gas velocity, pressure, density and temperature all rise 
to a small maximum. 
A weak shock occurs about midway in this compression zone. 
Interior to this is a rarefaction region of approximately uniform 
velocity in which the gas pressure, density and temperature all 
decline adiabatically. 
The gas temperature profiles show 
tiny peaks with subsequent cooling.
The flow velocities at all times shown are subsonic, 
$u < c_{s0} = 725$ km s$^{-1}$. 
The small (dynamically insignificant) 
oscillations in the velocity suggest that our 
modified artificial viscosity is not entirely damping out 
the kinetic energy entering the (weak) shock within the leading 
edge of the gas flow. 
At time $t = t_{lobe} = 10^7$ yrs, the cosmic rays continue 
to stream through the gas, providing a pressure 
gradient sufficient to
sustain the cavity in approximate hydrostatic equilibrium.

However, just after time $t_{lobe} = 10^7$ yrs 
when the CR luminosity ${\dot E}_c$ at the origin turns off, 
the CRs continue to diffuse rapidly through the gas. 
By time $t = 3 \times 10^7$ years shown in Figure 2 
the cosmic ray pressure has become much less than the 
gas pressure, allowing hot gas to flow back toward 
the center filling the cavity with nearly 
stationary gas, $u \approx 0$. 
The entropy near the center is seen to be slightly higher 
(lower density, higher temperature), indicating that 
this gas experienced a stronger shock when 
the CRs first turned on.

The upper panel of Figure 3 shows the evolution of the 
cavity radius, $r_{cav}(t)$, defined (arbitrarily) as that radius where 
the gas density drops to half its original value. 
The radius rises to a maximum $r_{cav} \approx 3$ kpc 
(about half $r_{lobe} = 6$ kpc) at $t = t_{lobe} = 10^7$ yrs, 
then drops catastrophically. 
Such a cavity cannot correspond to those observed which 
persist during their long buoyancy phase 
after the CR source has turned off.
We conclude that the diffusion coefficient in the gas must be less 
than $\kappa_0 = 10^{30}$ cm$^2$ s$^{-1}$ so that the cosmic 
rays can be confined within the cavity, sustaining 
it for a longer time.
The lower panel in Figure 3 shows the total CR energy that has 
entered the flow ${\dot E}_ct$ (solid line), 
the total CR energy in the computational grid 
$E_{c,tot}(t)$ (dashed-dotted line)
and the CR energy $E_{c,cav}(t)$ in the cavity 
$r < r_{cav}$ (dashed line). 

\section{Cavities with Density-dependent $\kappa$}

To prolong the observable lifetime of cavities 
beyond that of the previous calculation, we 
consider flows in which the cosmic ray diffusion is slower 
in the gas, but retains the value 
$\kappa_0 = 10^{30}$ cm$^2$ s$^{-1}$ in the cavity.
First we modify the diffusion coefficient so that 
$\kappa = {\tilde c}_{\kappa}\kappa_0$ where 
\begin{displaymath}
{\tilde c}_{\kappa} = g(n_e)~~~~~~~~~{\rm if}~~~ n_e > n_{ed}
\end{displaymath}
\begin{equation}
{\tilde c}_{\kappa} = 1~~~~~~~~~~~~~~~{\rm if}~~~ n_e < n_{ed}
\end{equation}
where 
\begin{displaymath}
g(n_e) = (n_e/n_{ed})^{-q},
\end{displaymath}
with $n_{ed} = 0.0063$ cm$^{-3}$ and $q = 1.48$. 
In the distant undisturbed gas 
${\tilde c}_{\kappa} = g(n_{e0}) = 0.10$ is about ten times lower 
than before.
However, during the initial stages of the calculation
when cosmic rays first encounter the gas with a low diffusion
coefficient near the origin,
the numerical solutions can become unreliable.
To avoid this difficulty, we also require that $\kappa$
have the value $\kappa_0$ near the origin regardless of the
gas density there.
To retain $c_{\kappa} \approx 1$ near the origin we 
adopt a final $\kappa = c_{\kappa}\kappa_0$ where 
\begin{equation}
c_{\kappa} = \case{1}{2}({\tilde c}_{\kappa} + 1 ) 
+ \case{1}{2}({\tilde c}_{\kappa} - 1)
\tanh \left[{(r-r_m) \over \Delta r_m} \right]
\end{equation}
and $r_m = \Delta r_m = 0.3$ kpc.
Once the cavity radius exceeds $\Delta r_m$, which occurs very 
early in the calculations, $\kappa = \kappa_0$ throughout the 
rest of the 
cavity and, as the gas density increases, $\kappa$ decreases as 
$\kappa_0g(n_e)$, approaching 
$\kappa_0g(n_{e0})$ in the distant undisturbed gas. 

Figures 4, 5, and 6 show the evolution of this cavity. 
At early times $t \lta 10^6$ yrs 
the stronger coupling of CRs with the gas results 
in mildly 
supersonic velocities and the maximum amplitudes of the flow
variables are somewhat greater than before. 
However, at later times $t > t_{lobe} = 10^7$ yrs (Figure 5) 
when the central CR source is no longer 
active, the cavity again closes rapidly, 
nearly returning to its initial state by $\sim 3 \times 10^7$ yrs.
The evolution of the cavity radius shown in Figure 6 
shows that the cavity produced at $10^7$ yrs is nearly twice 
as large as that in Figure 3. 
However, as before, the cavity radius 
drops immediately after time $t_{lobe}$, suggesting that 
our reduced $\kappa$ is still too large.

To further confine the cosmic rays, we show a final model 
similar to the previous one but with 
$n_{ed} = 0.0032$ cm$^{-3}$ and $q = 2.05$ so 
$g(n_{e0}) = (n_{e0}/n_{ed})^{-q} = 0.01$ and $\kappa \approx 10^{28}$ 
cm$^2$ s$^{-1}$ in the distant, undisturbed gas.
Figures 7, 8 and 9 show the corresponding cavity evolution. 
With this further reduction of CR diffusion in the gas,
the solution at times $t \lta t_{lobe}$ is similar to the previous 
calculation, but with even larger amplitudes in the flow 
variables. 
However, Figures 8 and 9 show that 
the cavity is now able to sustain a significant size after  
the CR source has been turned off at $t > t_{lobe} = 10^7$ yrs.
The flow profiles in Figure 8 at 
times $t = 3\times 10^7$ and $7 \times 10^7$ yrs 
illustrate this strong sustained confinement of
CRs within $r_{cav}$. 
The highly subsonic flow at these late times is nearly in hydrostatic 
equilibrium, but some low-density gas has flowed back into the cavity.
The upper panel of 
Figure 9 shows that the cavity radius reaches a maximum 
of $\sim 8$ kpc and by  
$3 \times 10^7$ years still has a radius $\sim 5$ kpc 
expected from the original choice of parameters. 

At times $t \lta t_{lobe} = 10^7$ yrs 
the CR energy in the cavity $E_{c,cav}$ and in the flow $E_{c,tot}$
(lower panel of Figure 9) both slightly 
exceed the total energy supplied ${\dot E}t$.
The total CR energy is not a strictly conserved quantity and 
since cavitation requires a net compression in the gas, 
cosmic ray energy can be created by the $e_c \propto \rho^{\gamma_c}$ 
term in Equation (9). 
However, at later times $t > t_{lobe} = 10^7$ yrs, 
when the cavity radius decreases, $E_{c,tot} \approx {\dot E}_ct$
to a good approximation.

The evolution of X-ray-radio cavities can also be seen in 
diffuse gamma 
ray emission either as inverse Compton (IC) emission 
from collisions of CR electrons 
with the cosmic microwave background or as 
the $\pi^0$ decay continuum resulting from CR proton 
collisions with hot gas protons. 
It is of interest to explore how the bolometric gamma ray 
emission due to $\pi^0$ decays varies with the evolution of 
X-ray cavities, assuming that protons are an important 
fraction of the total CR energy density.
If the spectral shape of the energy distribution of CR protons 
remains approximately constant with 
space and time, the number density of CR protons is 
proportional to $e_c$ and the local emissivity from $\pi^0$ decays 
is proportional to $e_c n_p$ where $n_p = 0.83 n_e$ is the 
proton density in the hot gas. 
Using the cavity evolution shown in Figures 7 and 8, 
Figure 10 shows the result of integrating 
$e_c n_p$  along the line of sight,
$\Sigma_{\gamma} = \int e_c n_p ds$, 
which should be approximately proportional to 
the radial surface brightness profile 
of the bolometric $\pi^0$ decay continuum. 

The dramatic evolution of the pion continuum 
shown in Figure 10 suggests that it may be possible to 
identify young cavities by a strong limb brightening in 
the gamma ray surface brightness, 
provided $\kappa \sim 10^{28}$ cm$^2$ s$^{-1}$.
However, if $\kappa < 10^{28}$ cm$^2$ s$^{-1}$ 
the pion continuum will be strongly peaked in the walls 
of cavities of all ages.
In general the shape of the CR diffusion front in gamma rays 
provides important information about the magnitude
of $\kappa$. 
Since the pion continuum due to proton CRs 
peaks at 70 Mev, it may be possible 
to detect the pion decay 
emission even in the presence of an appreciable IC continuum 
due to CR electrons. 
Moreover, the spectral difference in the two gamma ray 
continuua could in principle allow an estimation of the 
relative numbers of CR electrons and protons. 
Finally, the level of the pion continuum at the center 
of more mature cavities can provide information about 
the density of thermal gas within the cavities and 
therefore about the rate that cavity buoyancy is transporting 
mass outward in the hot atmosphere. 

\section{Final Remarks and Conclusions}

In these simple one-dimensional calculations we have
shown, perhaps for the first time, how radio lobes
and X-ray cavities are formed in galaxies and clusters.
We have not explored a vast range of parameters but have
concentrated instead on the critical influence of the 
CR diffusion coefficient $\kappa$ in cavity evolution.
Of particular interest is how the size and duration of the cavities 
depend on the rate that cosmic rays diffuse into the gas.
For the CR parameters adopted here, 
when the diffusion coefficient $\kappa$ is $\gta 10^{29}$
cm$^2$ s$^{-1}$,
the cavities tend to be smaller than those observed.
With these large values of $\kappa$
the cosmic rays diffuse readily through the cavity
walls into the hot gas, rapidly establishing a quasi-steady
cavity profile.
But these cavities also disappear rapidly following the
termination of the cosmic ray source.
This is unacceptable because cavities are observed during 
a long buoyancy time $t_{buoy}$ as they 
move radially away from their formation sites 
near the center of the group/cluster gas. 

However, when $\kappa \lta 10^{28}$ cm$^2$ s$^{-1}$ 
in the gas, with other parameters unchanged,
the cavities are larger, similar to those observed.
Furthermore, such cavities persist
$\sim 10$ times longer than the initial formation time,
an essential attribute of observed cavities.
It is noteworthy that $\kappa \sim 10^{28}$
cm$^2$ s$^{-1}$ is also the preferred diffusivity
in successful models of CR diffusion in the Milky Way 
(Jones, 1979;
Strong \& Moskalenko 1998;
Snodin, et al. 2006). 

Cavity formation by diffusing cosmic rays is particularly gentle
since it accelerates the ambient gas
over an extended region of thickness 
$\sim (\kappa t)^{1/2}$,
not at a surface as occurs 
when cosmic rays are approximated with a hot gas. 
As a result the shocks produced by CRs in the surrounding
gas are mild and rather insignificant in our
calculations.
To simulate the varying density of magnetic CR scatterers 
that scale with the gas density, 
we consider solutions in which the 
diffusion coefficient varies inversely with some power of 
the gas density.
Finally, the long-lived cavities we describe here 
are formed in a hot gas with uniform 
pressure and are therefore expected to be somewhat smaller 
than observed cavities which expand as they 
experience lower ambient 
pressures during their buoyant outward motion 
in the hot gas atmosphere of a galaxy cluster.

Cosmic rays are confined by partial 
reflection from the cavity walls.
This point is of considerable interest since
radio observations of X-ray cavities also 
indicate that synchrotron emitting electrons are
confined within X-ray cavities
(e.g. for Hydra A: McNamara et al. 2000; David et al. 2001;
Allen et al 2001).
Furthermore, proton CRs with energies exceeding
the $\pi^{0}$ rest mass (135 Mev), if trapped in cavities,
will experience fewer collisions in the low-density
cavity plasma, resulting in lower global gamma ray luminosities 
from $\pi^0 \rightarrow 2\gamma$ decays,
as indicated by
EGRET observations (e.g. Pfrommer \& Ensslin 2004),
than if the CRs were uniformly distributed throughout the 
hot gas as often assumed.
Figure 10 shows that $\pi^0$ production is enhanced 
in the dense gaseous walls of young cavities 
and in the cores of old cavities.
Observations of the $\pi^{0} \rightarrow \gamma$ decay continuum 
are expected to provide 
important information about the cavity age, 
the CR diffusivity, the relative numbers of CR electrons and protons 
and the density and mass of thermal gas within the cavity.
At 100 Mev GLAST is expected to have a spatial resolution
of 30$^{\prime \prime}$ which may be sufficient to
resolve gamma ray profiles in the pion continuum 
in nearby groups/clusters.
Inverse Compton emission from radio lobes
is also expected to be sensitive to $\kappa$. 

Finally, we note that during the formation of young cavities
the cosmic ray pressure $P_c$ inside the cavity can exceed
the external gas pressure by factors of 3-30 
(e.g. Fig. 7) 
while studies of the radio emission from the
cavity have traditionally assumed equal pressures within
the cavity and the ambient gas
(e.g. Dunn \& Fabian 2004).

\vskip.1in

Studies of the evolution of hot gas in elliptical galaxies
at UC Santa Cruz are supported by
NASA grants NAG 5-8409 \& ATP02-0122-0079 and NSF grant
AST-0098351 for which we are very grateful.
Continued support would also be appreciated.


\clearpage
\begin{figure}
\centering
\vskip2.in
\includegraphics[bb=90 166 522 519,scale=0.8,angle= 270]
{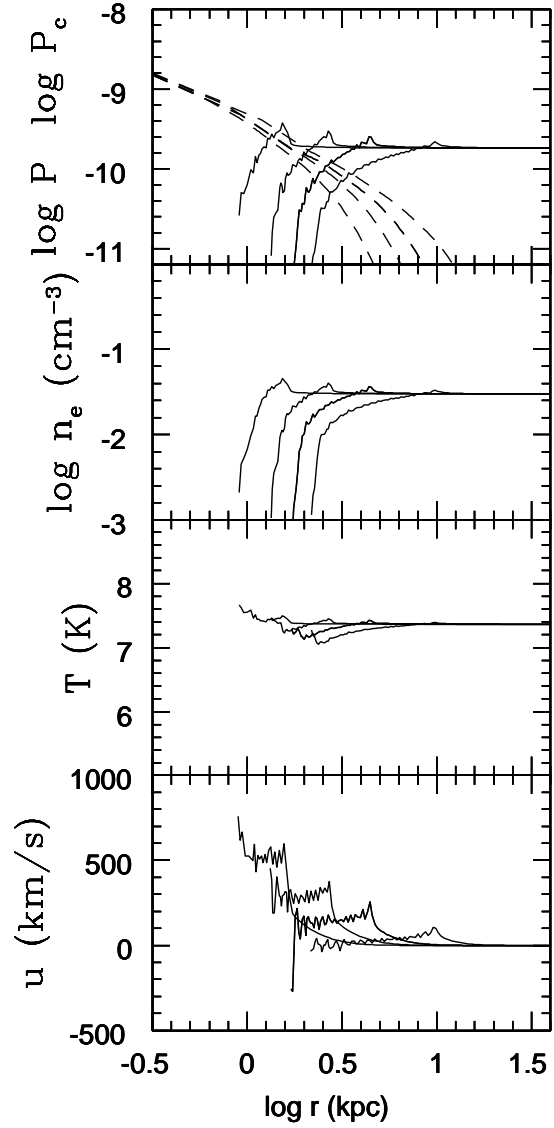}
\vskip.7in
\caption{
Flow variables for cavity evolution with
constant cosmic ray diffusion coefficient 
$\kappa = \kappa_0 = 1.08 \times 10^{30}$ cm$^2$ s$^{-1}$.
The evolution is shown at four times 
0.080, 0.183, 0.611 and $1.000\times 10^7$ yrs 
and all profiles evolve 
toward larger radii as the cavity forms.
In the top panel the dashed lines show the cosmic 
ray pressure and the solid lines show the gas pressure. 
}
\label{f1}
\end{figure}

\clearpage
\begin{figure}
\vskip2.in
\centering
\includegraphics[bb=90 166 522 519,scale=0.8,angle= 270]
{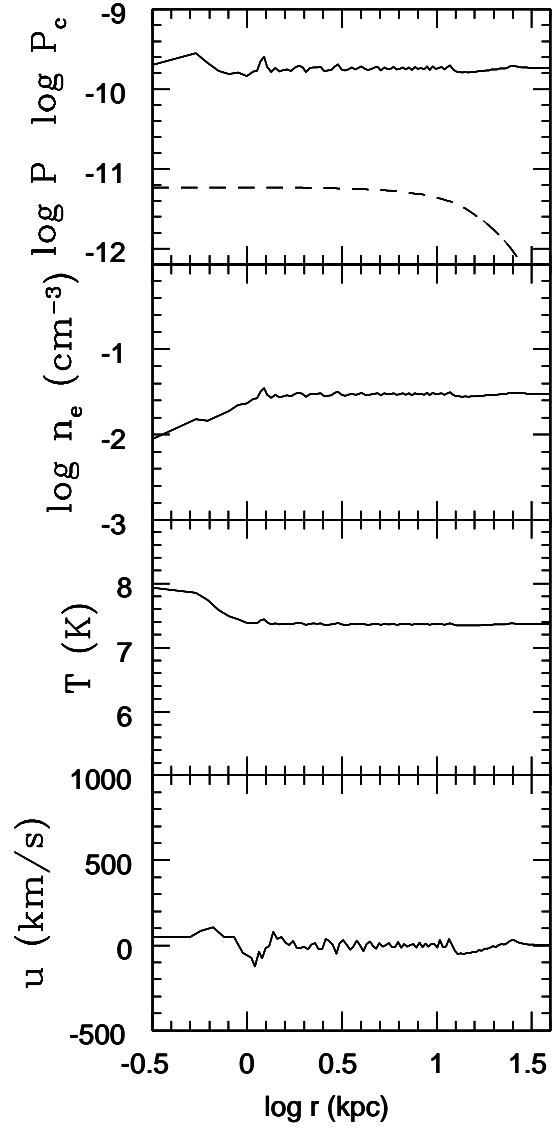}
\vskip.7in
\caption{
Flow variables for cavity evolution with
constant cosmic ray diffusion coefficient
$\kappa = \kappa_0 = 1.08 \times 10^{30}$ cm$^2$ s$^{-1}$.
The evolution is shown at time $t = 3.0 \times 10^7$ yrs. 
In the top panel the dashed line shows the cosmic
ray pressure and the solid line shows the gas pressure.
}
\label{f2}
\end{figure}

\clearpage
\begin{figure}
\vskip2.in
\centering
\includegraphics[bb=90 166 522 519,scale=0.8,angle= 270]
{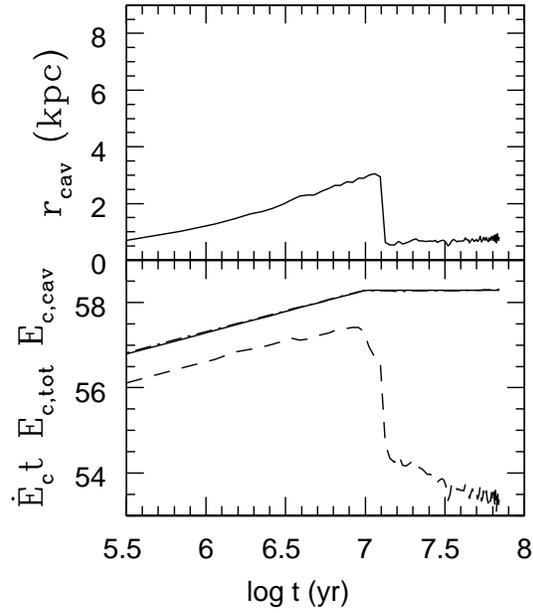}
\vskip.7in
\caption{
Evolution of cavity with constant cosmic ray diffusion coefficient.
{\it Top panel}: Evolution of the radius of the cavity 
$r_{cav}(t)$.
{\it Lower panel}: evolution of the total CR energy that has
entered the flow ${\dot E}t$ (solid line),
the total cosmic ray energy in the computational grid
$E_{c,tot}$ (dashed-dotted line)
and the CR energy $E_{c,cav}$ in the cavity
$r < r_{cav}$ (dashed line).
The variation of $E_{c,tot}$ closely follows ${\dot E}_ct$.
}
\label{f3}
\end{figure}

\clearpage
\begin{figure}
\vskip2.in
\centering
\includegraphics[bb=90 166 522 519,scale=0.8,angle= 270]
{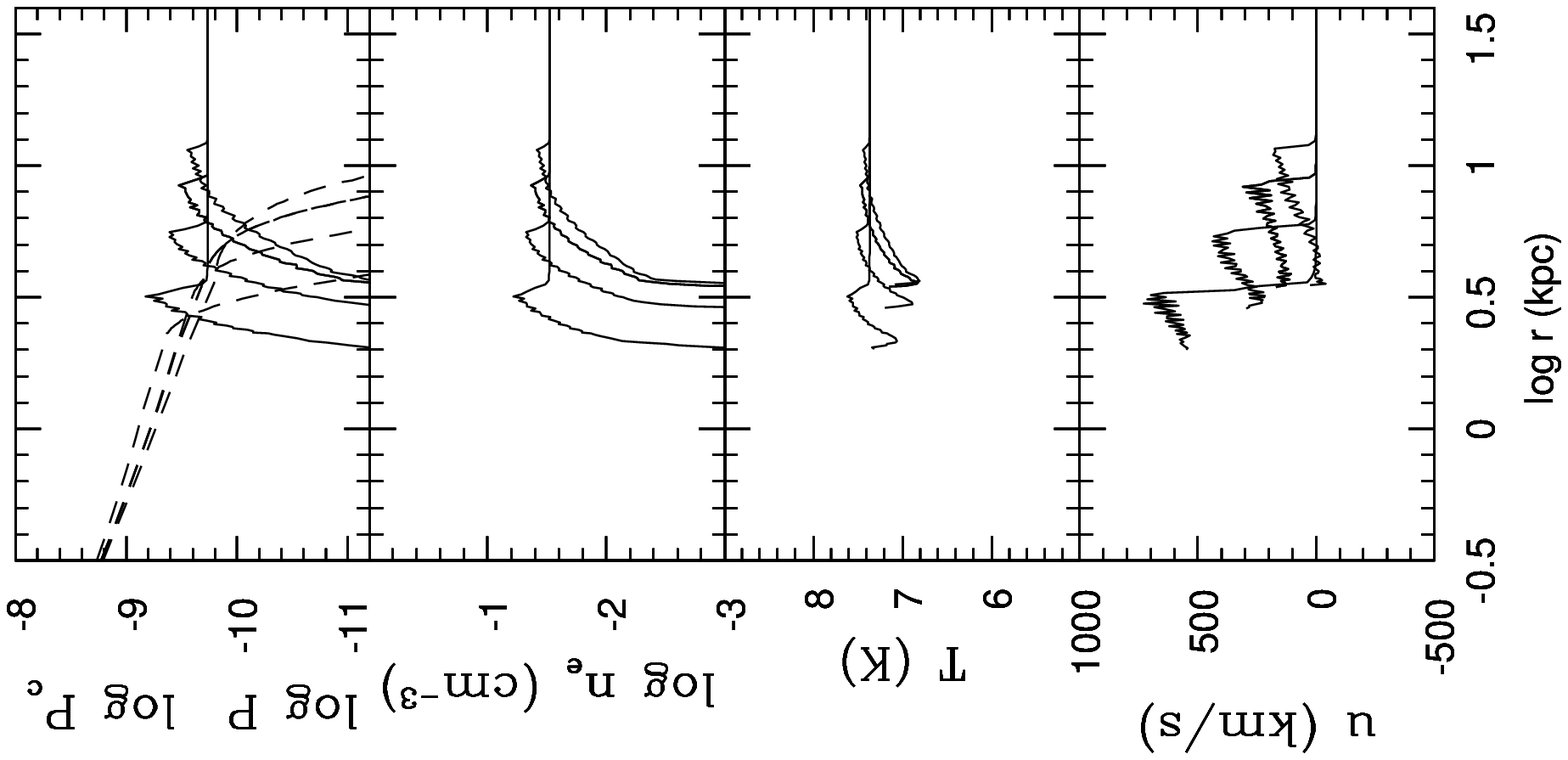}
\vskip.7in
\caption{
Flow variables for cavity evolution with 
density-dependent cosmic ray diffusion.
The diffusion coefficient in the cavity
$\kappa = \kappa_0 = 1.08 \times 10^{30}$ cm$^2$ s$^{-1}$
is constant but in the gas the diffusion coefficent increases
as $\kappa \propto (n_{e}/0.00631)^{-1.477}$, reaching 
$\kappa = 0.1\kappa_0$ in the distant gas where $n_e = n_{e0}$.
The evolution is shown at four times
0.169, 0.368, 0.668 and $1.000\times 10^7$ yrs. 
All profiles evolve
toward larger radii as the cavity forms.
In the top panel the dashed lines show the cosmic
ray pressure and the solid lines show the gas pressure.
}
\label{f4}
\end{figure}

\clearpage
\begin{figure}
\vskip2.in
\centering
\includegraphics[bb=90 166 522 519,scale=0.8,angle= 270]
{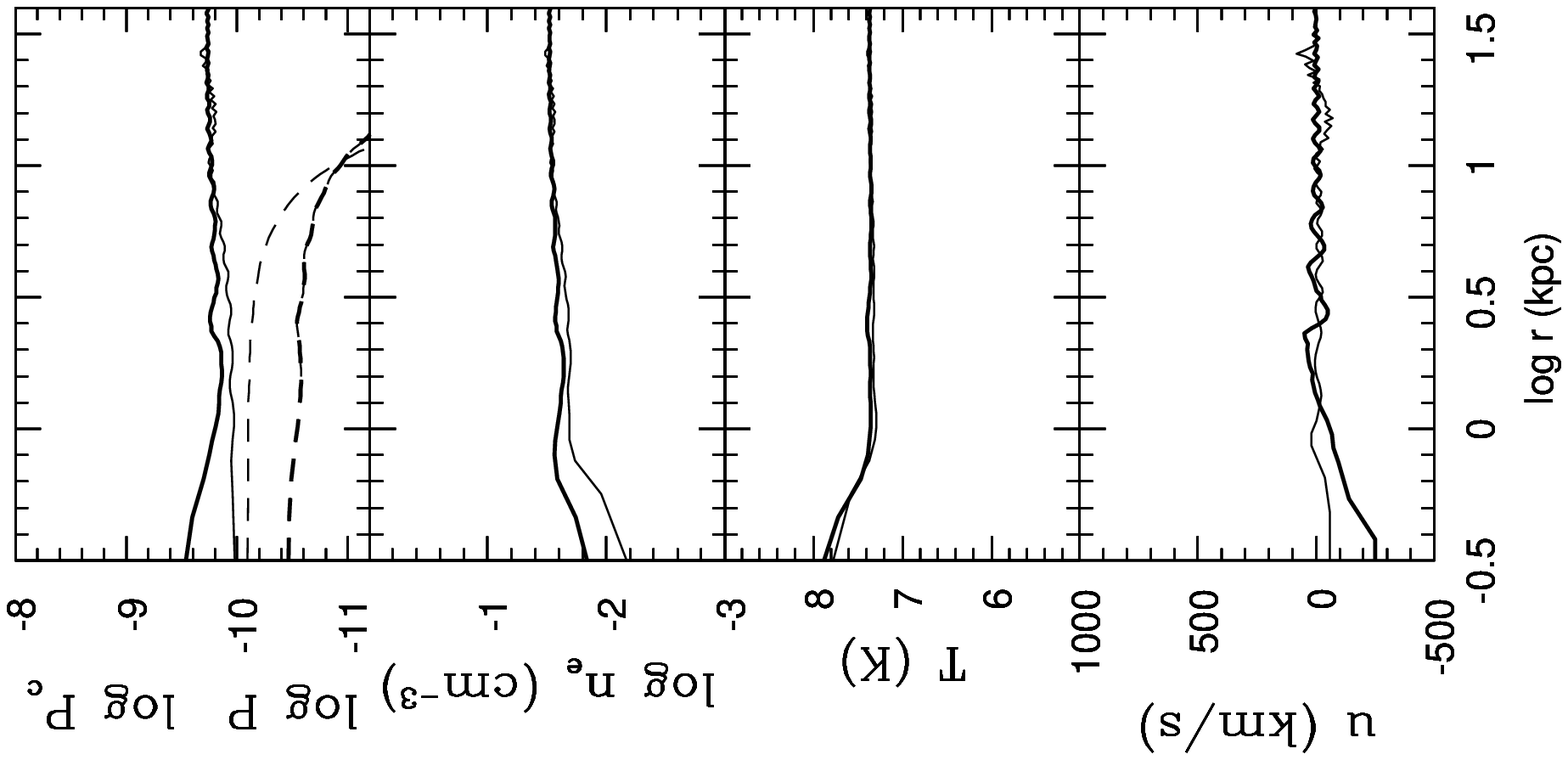}
\vskip.7in
\caption{
Flow variables for cavity evolution with
density-dependent cosmic ray diffusion.
The diffusion coefficient in the cavity
$\kappa = \kappa_0 = 1.08 \times 10^{30}$ cm$^2$ s$^{-1}$
is constant but in the gas the diffusion coefficent increases
as $\kappa \propto (n_{e}/0.00631)^{-1.477}$, reaching
$\kappa = 0.1\kappa_0$ in the distant gas where $n_e = n_{e0}$.
The flow profiles are shown at two times
$3.00 \times 10^7$ yrs (light lines) and  
$7.00 \times 10^7$ yrs (heavy lines). 
In the top panel the dashed lines show the cosmic
ray pressure and the solid lines show the gas pressure.
}
\label{f5}
\end{figure}

\clearpage
\begin{figure}
\vskip2.in
\centering
\includegraphics[bb=90 166 522 519,scale=0.8,angle= 270]
{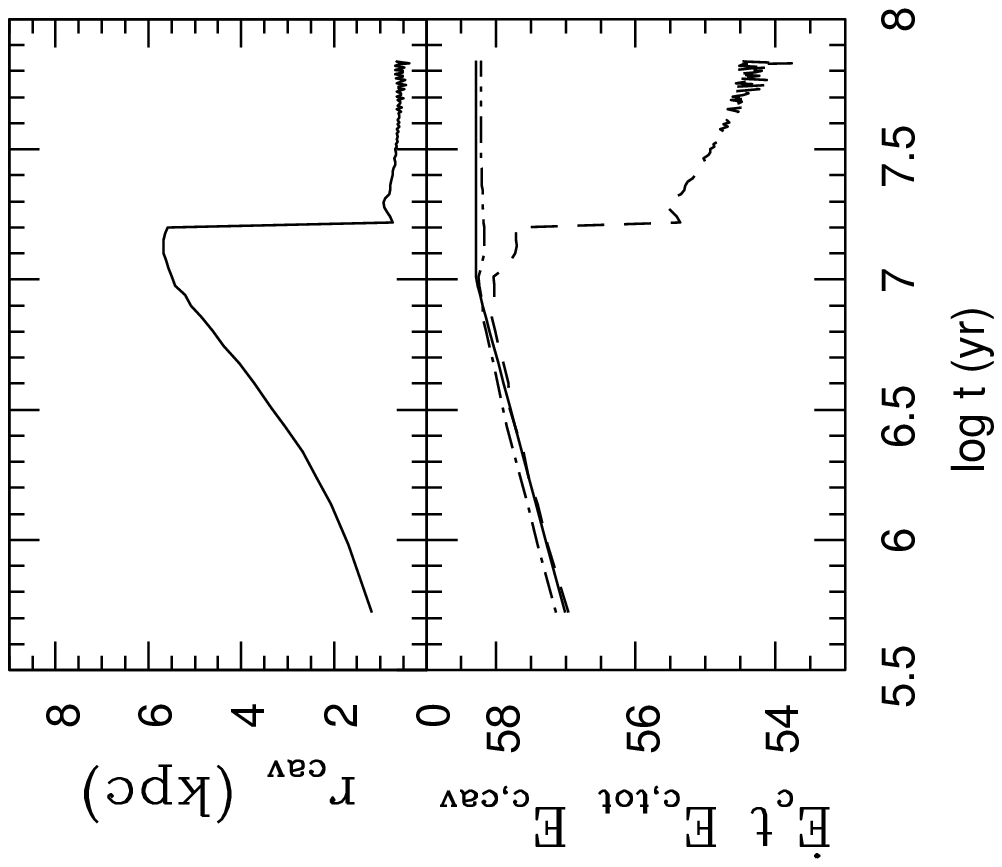}
\vskip.7in
\caption{
Evolution of cavity with density-dependent 
cosmic ray diffusion coefficient, reaching 
$\kappa = 0.1\kappa_0$ in the distant undisturbed gas. 
{\it Top panel}: Evolution of the radius of the cavity
$r_{cav}(t)$.
{\it Lower panel}: evolution of the total CR energy that has
entered the flow ${\dot E}t$ (solid line),
the total cosmic ray energy in the computational grid
$E_{c,tot}$ (dashed-dotted line)
and the CR energy $E_{c,cav}$ in the cavity
$r < r_{cav}$ (dashed line).
}
\label{f6}
\end{figure}

\clearpage
\begin{figure}
\vskip2.in
\centering
\includegraphics[bb=90 166 522 519,scale=0.8,angle= 270]
{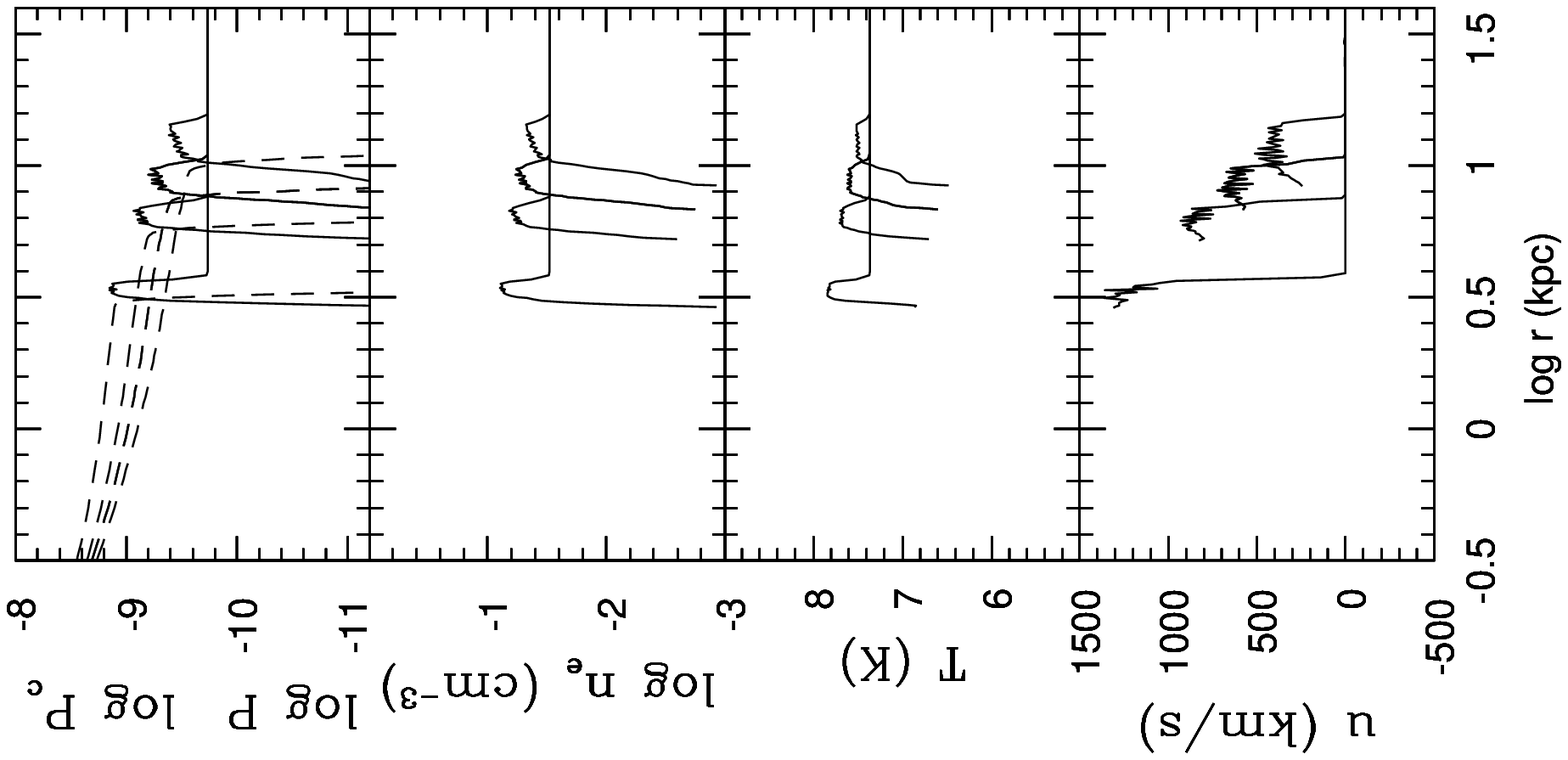}
\vskip2.in
\caption{
Flow variables for cavity evolution with
density-dependent cosmic ray diffusion.
The diffusion coefficient in the cavity
$\kappa = \kappa_0 = 1.08 \times 10^{30}$ cm$^2$ s$^{-1}$
is constant but in the gas the diffusion coefficient increases
as $\kappa \propto (n_{e}/0.00316)^{-2.047}$, reaching
$\kappa = 0.01\kappa_0$ in the distant gas where $n_e = n_{e0}$.
The evolution is shown at four times
0.156, 0.375, 0.600 and $1.000\times 10^7$ yrs.
All profiles evolve
toward larger radii as the cavity forms.
In the top panel the dashed lines show the cosmic
ray pressure and the solid lines show the gas pressure.
}
\label{f7}
\end{figure}

\clearpage
\begin{figure}
\vskip2.in
\centering
\includegraphics[bb=90 166 522 519,scale=0.8,angle= 270]
{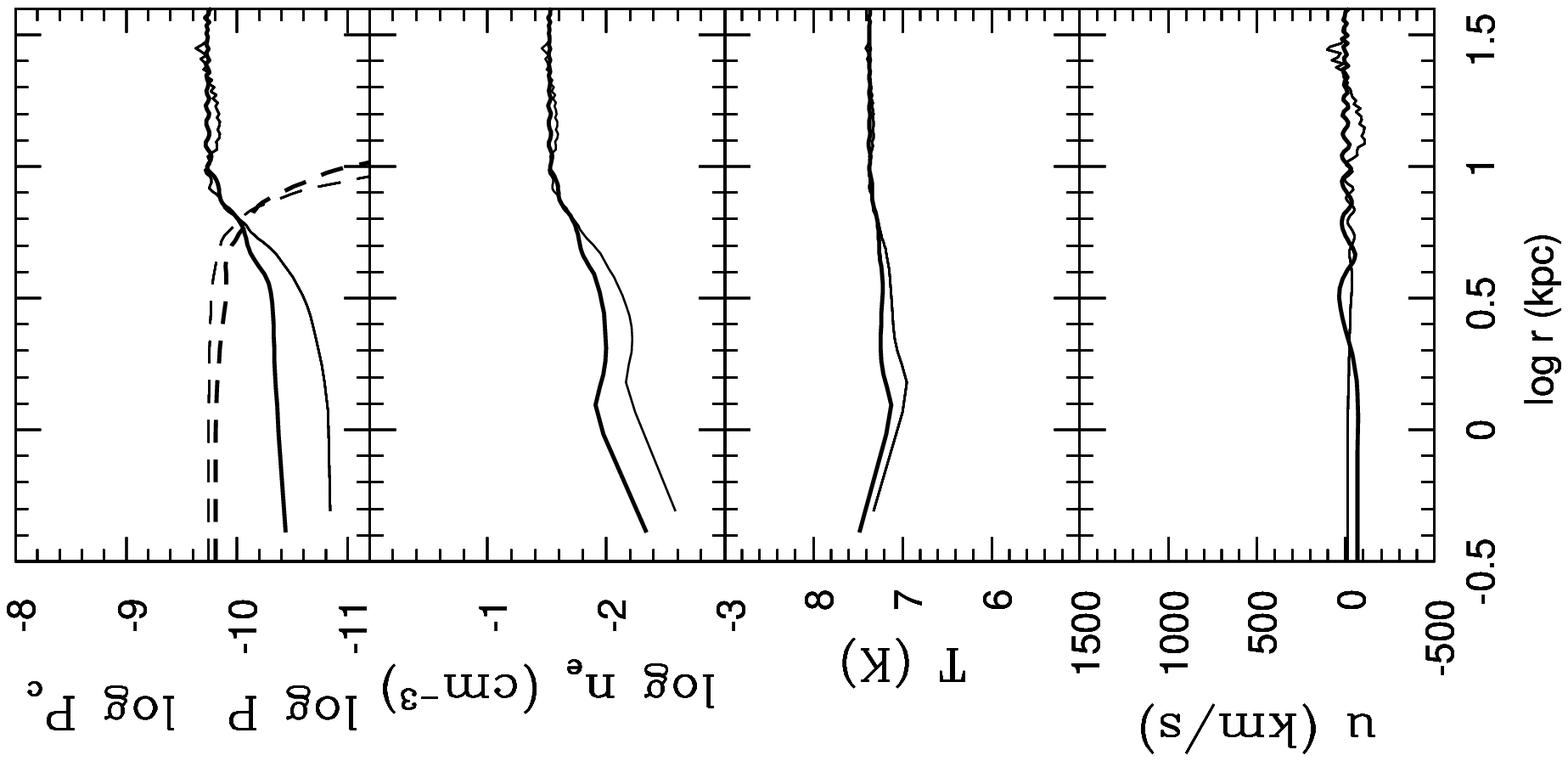}
\vskip.7in
\caption{
Flow variables for cavity evolution with
density-dependent cosmic ray diffusion.
The diffusion coefficient in the cavity
$\kappa = \kappa_0 = 1.08 \times 10^{30}$ cm$^2$ s$^{-1}$
is constant but in the gas the diffusion coefficient increases
as $\kappa \propto (n_{e}/0.00316)^{-2.047}$, reaching
$\kappa = 0.01\kappa_0$ in the distant gas where $n_e = n_{e0}$.
The flow profiles are shown at two times
$3.00 \times 10^7$ yrs (light lines) and
$7.00 \times 10^7$ yrs (heavy lines).
In the top panel the dashed lines show the cosmic
ray pressure and the solid lines show the gas pressure.
}
\label{f8}
\end{figure}

\clearpage
\begin{figure}
\vskip2.in
\centering
\includegraphics[bb=90 166 522 519,scale=0.8,angle= 270]
{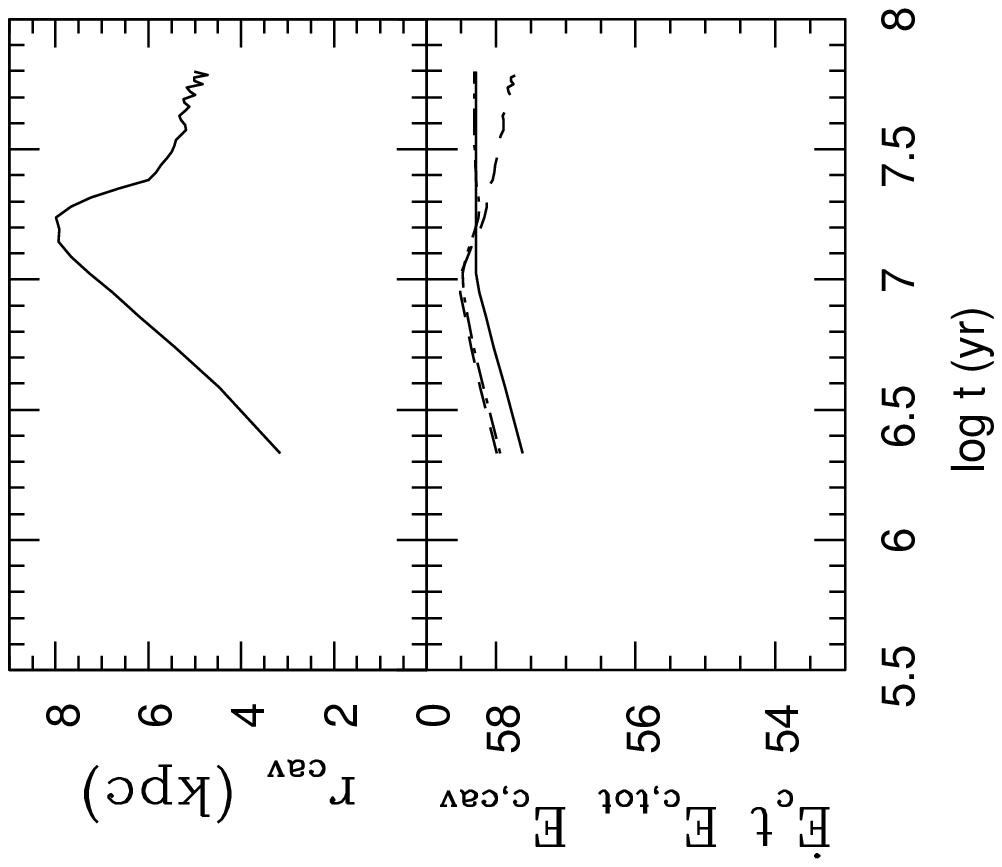}
\vskip.7in
\caption{
Evolution of cavity with density-dependent
cosmic ray diffusion coefficient, reaching
$\kappa = 0.01\kappa_0$ in the distant undisturbed gas.
{\it Top panel}: Evolution of the radius of the cavity
$r_{cav}(t)$.
{\it Lower panel}: evolution of the total CR energy that has
entered the flow ${\dot E}t$ (solid line),
the total cosmic ray energy in the computational grid
$E_{c,tot}$ (dashed-dotted line)
and the CR energy $E_{c,cav}$ in the cavity
$r < r_{cav}$ (dashed line).
The variation of $E_{c,tot}(t)$ closely follows ${\dot E}_ct$
at $t \lesssim 10^7$ years, and closely follows 
${\dot E}_ct$ thereafter.
}
\label{f9}
\end{figure}

\clearpage
\begin{figure}
\vskip2.in
\centering
\includegraphics[bb=90 166 522 519,scale=0.8,angle= 270]
{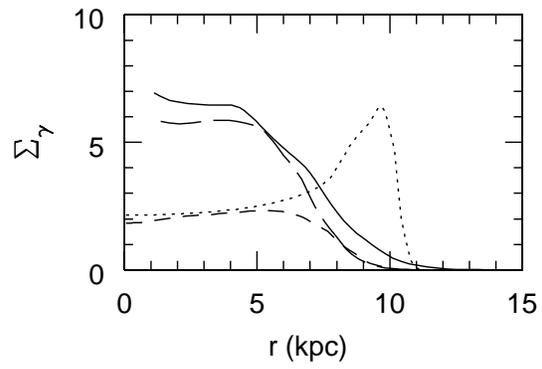}
\vskip.7in
\caption{
Approximate variation of the bolometric 
pion gamma ray continuum for 
the cavity shown in Figures 7 and 8. 
The gamma ray surface brightness 
$\Sigma_{\gamma}$ is shown in arbitrary units at four 
times: $1.0 \times 10^7$ yrs (dotted line), 
$1.9 \times 10^7$ yrs (short dashed line),
$1.0 \times 10^7$ yrs (long dashed line),
and $7.0 \times 10^7$ yrs (solid line).
}
\label{f10}
\end{figure}

\end{document}